\documentstyle[preprint,aps,epsf]{revtex}                                   
\begin{document}
\newcommand{\be}{\begin{equation}}
\newcommand{\ee}{\end{equation}}
\newcommand{\bea}{\begin{eqnarray}}
\newcommand{\eea}{\end{eqnarray}}

\title{Density of states, Potts zeros, and
Fisher zeros of the $Q$-state Potts model for continuous $Q$} 
\author{Seung-Yeon Kim
\footnote{Current address: Department of Chemical Engineering,
Princeton University, Princeton, New Jersey 08544.
Electronic address: seungk@princeton.edu}
and Richard J. Creswick\footnote{Electronic address: creswick.rj@sc.edu}}
\address{Department of Physics and Astronomy,
University of South Carolina,\\ Columbia, South Carolina 29208}
\maketitle

\begin{abstract}
The $Q$-state Potts model can be extended to non-integer and even 
complex $Q$ by expressing the partition function in the Fortuin-Kasteleyn
(F-K) representation. In the F-K representation the partition function,
$Z(Q,a)$, is a polynomial in $Q$ and $v=a-1$ ($a=e^{\beta J}$) and the 
coefficients of this polynomial, $\Phi(b,c)$, are the number of graphs on 
the lattice consisting of $b$ bonds and $c$ connected clusters. We introduce 
the random-cluster transfer matrix to compute $\Phi(b,c)$ exactly on 
finite square lattices with several types of boundary conditions.
Given the F-K representation of the partition function we begin by
studying the {\it critical Potts model} $Z_{CP}=Z(Q,a_c(Q))$,
where $a_c(Q)=1+\sqrt{Q}$. We find a set of zeros in the complex 
$w=\sqrt{Q}$ plane that map to (or close to) the Beraha numbers for real 
positive $Q$. We also identify $\tilde{Q}_c(L)$, the value of $Q$ for
a lattice of width $L$ above which the locus of zeros in the complex
$p=v/\sqrt{Q}$ plane lies on the unit circle. By finite-size scaling
we find that $1/\tilde{Q}_c(L)\to0$ as $L\to\infty$. We then study 
zeros of the antiferromagnetic (AF) Potts model in the complex $Q$ plane
and determine $Q_c(a)$, the largest value of $Q$ for a fixed value of
$a$ below which there is AF order. We find excellent agreement with
Baxter's conjecture $Q_c^{AF}(a)=(1-a)(a+3)$.
We also investigate the locus of zeros of the ferromagnetic Potts model
in the complex $Q$ plane and confirm that $Q_c^{FM}(a)=(a-1)^2$. We show
that the edge singularity in the complex $Q$ plane approaches $Q_c$ as
$Q_c(L)\sim Q_c+AL^{-y_q}$, and determine the scaling exponent $y_q$ for
several values of $Q$. Finally,   
by finite size scaling of the Fisher zeros near the antiferromagnetic
critical point we determine the thermal exponent $y_t$ as a function of
$Q$ in the range $2\le Q\le3$. Using data for lattices of size $3\le L\le8$
we find that $y_t$ is a smooth function of $Q$ and is well fit by 
$y_t={1+Au+Bu^2\over C+Du}$ where $u=-{2\over\pi}\cos^{-1}{\sqrt{Q}\over2}$. 
For $Q=3$ we find $y_t\simeq0.6$; however if we include 
lattices up to $L=12$ we find $y_t\simeq0.50(8)$ in rough agreement with
a recent result of Ferreira and Sokal 
$[$J. Stat. Phys. {\bf96}, 461 (1999)$]$.   
\end{abstract}

\vskip2cm
PACS number(s): 05.10.$-$a, 05.50.+q, 64.60.Cn, 75.10.Hk


\section{introduction}

The $Q$-state Potts model\cite{wu} in two dimensions
exhibits a rich variety of critical behavior
and is very fertile ground for the analytical and numerical
investigation of first- and second-order phase transitions.
With the exception of the $Q=2$ Potts (Ising) model
in the absence of an external magnetic field, 
exact solutions for arbitrary $Q$ are not known.
However, some exact results at the critical temperature
have been established for the $Q$-state Potts model.
From the duality relation the ferromagnetic critical temperature
is known to be $T_c=J/k_B{\rm ln}(1+\sqrt{Q})$ for the 
isotropic square lattice.
Baxter\cite{baxter1} calculated the free energy at $T_c$ in the 
thermodynamic limit, and showed that the Potts model has 
a second-order phase transition for $Q\le4$ and a first-order 
transition for $Q>4$. 
The critical exponents for the ferromagnetic Potts 
model are well known\cite{nienhuis,pearson,blote}.  

On the other hand, the antiferromagnetic Potts model is much less well
understood than the ferromagnetic model.
Recently the three-state Potts antiferromagnet on the square lattice
has attracted a good deal of interest\cite
{grest,cardy,jayaprakash,nightingale,baxter2,dennijs,temesvari,fucito,racz,kolafa,wang,park,saleur,bakchich,sokal1,burton,sokal2,shrock1,sokal3,moore,shrock2}.  
Baxter\cite{baxter2} conjectured that the critical point
of the Potts antiferromagnet on the square lattice is given by
$T_c=J/k_B{\rm ln}(\sqrt{4-Q}-1)$, and
evaluated the critical free energy and internal energy.
The Baxter formula for the critical temperature gives 
the known exact value for $Q=2$,
a critical point at zero temperature for $Q=3$, 
and no critical point for $Q>3$.
For continuous $Q$ in the range $0<Q<3$, Kim {\it et al.}\cite{kim1} 
have studied
the antiferromagnetic Potts critical point through the zeros of the
partition function and found good agreement with the Baxter formula.   
With the exception of the Ising model,
the critical exponents of the Potts antiferromagnets are not known. 
However, for $Q=3$ the ratio of critical exponents
$\gamma/\nu$ is known to be 5/3\cite{wang,park}.

By introducing the concept of the zeros of the partition function
in the {\it complex} magnetic-field plane (Yang-Lee zeros),
Yang and Lee\cite{yang} proposed a mechanism
for the occurrence of phase transitions in the thermodynamic limit
and yielded a new insight into the unsolved problem of the Ising model
in an arbitrary nonzero external magnetic field. 
It has been shown\cite{yang,creswick1,kenna1} 
that the distribution of the zeros of a model 
determines its critical behavior. Lee and Yang\cite{yang} also 
formulated the celebrated circle theorem which states that
the Yang-Lee zeros of the Ising ferromagnet lie on the unit circle
in the complex magnetic-field ($x=e^{\beta h}$) plane. 
However, for the $Q$-state Potts model with $Q>2$ the Yang-Lee zeros 
lie close to, but not on, the unit circle with the two exceptions of 
the critical point $x=1$ ($h=0$) itself and the  
zeros in the limit $T=0$\cite{kim2}.

Fisher\cite{fisher} emphasized that the partition
function zeros in the complex temperature plane (Fisher zeros)
are also very useful in understanding phase transitions, and
showed that for the square lattice Ising model in the absence of 
an external magnetic field the Fisher zeros
lie on two circles in the thermodynamic limit.
In particular, using the Fisher zeros both the ferromagnetic
phase and the antiferromagnetic phase can be considered
at the same time.
The critical behavior of the Potts model in both
the ferromagnetic and antiferromagnetic phases
have been studied using the distribution of the Fisher zeros, and 
the Baxter conjecture for the antiferromagnetic critical temperature
has been verified\cite{kim1}. 
Recently the Fisher zeros of the $Q$-state Potts model on square 
lattices have been studied extensively for integer $Q>2$\cite
{maillard,martin1,martin2,martin3,wood,bhanot,alves,martin4,chen1,chen2,shrock3,creswick2,kenna2,kim3} 
and noninteger $Q$\cite{kim1}. 
Exact numerical studies have shown\cite
{kim1,martin2,martin3,martin4,chen1,shrock3,creswick2,kim3} that
for self-dual boundary conditions the Fisher zeros of the $Q>1$ Potts models
on a finite square lattice are located on the unit circle in the complex
$p$ plane for ${\rm Re}(p)>0$, where $p=(e^{\beta J}-1)/\sqrt{Q}$.
It has been analytically shown that all the Fisher zeros of the 
infinite-state Potts model lie on the unit circle for any size of
square lattice with self-dual boundary conditions\cite{chen2}, and
the Fisher zeros near the ferromagnetic critical point of the 
$Q>4$ Potts models on the square lattice lie on the unit circle
in the thermodynamic limit\cite{kenna2}.   
Chen {\it et al.}\cite{chen1} conjectured that when $Q$ reaches a certain 
critical value $\tilde{Q}_c(L)$, all Fisher zeros for $L\times L$ square
lattices with self-dual boundary conditions are located at the unit 
circle $|p|=1$. In this paper we verify this conjecture and find that 
$\tilde{Q}_c(L)$ approaches infinity in the thermodynamic limit,
and we study the thermal exponent $y_t$ of the 
square-lattice Potts antiferromagnet 
using the Fisher zeros near the antiferromagnetic critical point.    
  
In this paper we also discuss the partition 
function zeros in the complex $Q$ plane (Potts zeros) of the 
$Q$-state Potts model. The Potts zeros at $\beta J=-\infty$ have
been investigated extensively to understand the ground states
of the antiferromagnetic Potts model and the chromatic polynomial
in graph theory\cite{shrock1,shrock2,baxter3,shrock4,biggs,shrock5,sokal4}. 
Recently the Potts zeros at finite temperatures have been studied for 
cyclic ladder graphs and $e^{\beta J}\le1$\cite{shrock5}.  

In the next section we describe two algorithms to evaluate the density
of states, from which the exact partition function of the $Q$-state 
Potts model is obtained.
The first algorithm (microcanonical transfer matrix) is applied to 
only integer $Q$ but allows us 
to calculate the density of states for relatively larger lattices,
while the second algorithm (random-cluster transfer matrix) 
gives the density of states for any value of $Q$.
In Sec. III we discuss the Potts model at $a=e^{\beta J}=1\pm\sqrt{Q}$,
its Potts zeros, and the related properties of the Fisher zeros.
In the subsequent two sections we study the Potts zeros for the 
antiferromagnetic interval $0\le a\le1$ (Sec. IV) and for the 
ferromagnetic interval $a\ge1$ (Sec. V).
In Sec. VI we discuss the thermal exponent $y_t$ of the 
square lattice $Q$-state Potts antiferromagnet for $2\le Q\le3$
using the Fisher zeros.    


\section{density of states}

The $Q$-state Potts model for integer $Q$
on a lattice $G$ with  $N_s$ sites and $N_b$ bonds 
is defined by the Hamiltonian
\be
{\cal H}_Q=-J\sum_{\langle i,j\rangle} \delta(\sigma_i,\sigma_j),
\ee
where $J$ is the coupling constant, 
$\langle i,j\rangle$ indicates a sum over nearest-neighbor pairs,
$\delta$ is the Kronecker delta,
and $\sigma_i=1,2,...,Q$.
The partition function of the model is
\be
Z_Q=\sum_{\{ \sigma_n \}} e^{-\beta{\cal H}_Q},
\ee
where $\{ \sigma_n\}$ denotes a sum over $Q^{N_s}$ possible spin 
configurations and $\beta=(k_B T)^{-1}$.
If we define the density of states with energy $0\le E\le N_b$ by 
\be
\Omega_Q(E)=\sum_{\{ \sigma_n \} }
\delta(E-\sum_{\langle i,j\rangle} \delta(\sigma_i,\sigma_j)),
\ee
which takes on only integer values,
then the partition function can be written as 
\be 
Z_Q(a)=\sum_{E=0}^{N_b}\Omega_Q(E) a^E,
\ee
where $a=e^{\beta J}$
and states with $E=0$ ($E=N_b$) correspond to the antiferromagnetic 
(ferromagnetic) ground states. From Eq. (4) it is clear that $Z_Q(a)$
is simply a polynomial in $a$. We have calculated exact integer values
for $\Omega_{Q=3}(E)$ of the three-state Potts model on finite 
$L\times L$ square lattices up to $L=12$ 
using the microcanonical transfer matrix ($\mu$TM)\cite{creswick3}.

Here we describe briefly the $\mu$TM\cite{creswick3} 
on an $L\times N$ square lattice 
with periodic boundary conditions in the horizontal direction (length $L$) 
and free boundaries in the vertical direction (length $N$).
First, an array, $\omega^{(1)}$, which is indexed by the energy $E$ 
and variables $\sigma_i$, $1\le i\le L$ for the first row
of sites is initialized as
\be
\omega^{(1)}(E;\sigma_1,\sigma_2,...,\sigma_L)=
\delta(E-\sum_{i=1}^L \delta(\sigma_i,\sigma_{i+1})).
\ee
Now each spin in the row is traced over in turn, 
introducing a new spin variable from the next row,
\be
\tilde{\omega}(E;\sigma'_1,\sigma_2,...,\sigma_L)=
\sum_{\sigma_1}\omega^{(1)}(E-\delta(\sigma'_1,\sigma_1);
\sigma_1,\sigma_2,...,\sigma_L).
\ee
This procedure is repeated until all the spins in the first row
have been traced over, leaving a new function of the $L$ spins
in the second row. The horizontal bonds connecting the spins 
in the second row are then taken into account by shifting
the energy,
\be
\omega^{(2)}(E;\sigma'_1,\sigma'_2,...,\sigma'_L)=
\tilde{\omega}(E-\sum_{i=1}^L\delta(\sigma'_i,\sigma'_{i+1});
\sigma'_1,\sigma'_2,...,\sigma'_L).
\ee
This procedure is then applied to each row in turn until the final ($N$th)
row is reached. The density of states is then given by
\be
\Omega_Q(E)=\sum_{\sigma'_1}\sum_{\sigma'_2}...\sum_{\sigma'_L}
\omega^{(N)}(E;\sigma'_1,\sigma'_2,...,\sigma'_L).
\ee 
The permutation symmetry of the $Q$-state Potts model allows us to
freeze the last spin $\sigma_L=1$ of each row.
Now we need to consider only $Q^{L-1}$ possible spin configurations 
in each row instead of $Q^L$ configurations, and we save a great amount
of memory and CPU time. 

On the other hand, Fortuin and Kasteleyn\cite{fortuin} have shown that 
the partition function is also given by 
\be
Z(a,Q)=\sum_{G^\prime\subseteq G} (a-1)^{b(G^\prime)} Q^{c(G^\prime)},
\ee
where the summation is taken over all subgraphs $G^\prime \subseteq G$,
and $b(G^\prime)$ and $c(G^\prime)$ are, respectively, the number of 
occupied bonds and clusters in $G^\prime$.
In Eq. (9) $Q$ need not be an integer and Eq. (9) defines 
the partition function of the $Q$-state Potts model for continuous $Q$.
The random-cluster (or Fortuin-Kasteleyn) representation of 
the Potts model, Eq. (9), is
also known as the Tutte dichromatic polynomial or the Whitney rank 
function in graph theory\cite{shrock5,sokal4}.
Introducing the density of states indexed by the number of occupied
bonds $0\le b\le N_b$ and the number of clusters $1\le c\le N_s$,   
\be
\Phi(b,c)=\sum_{G'}\delta(b-b(G'))\delta(c-c(G')),
\ee
which also takes on only integer values, the random-cluster 
representation of the Potts model can be written as 
\be
Z(a,Q)=\sum_{b=0}^{N_b}\sum_{c=1}^{N_s}\Phi(b,c)(a-1)^b Q^c,
\ee
which is again a polynomial in $a-1$ and $Q$.
We have evaluated exact integer values for $\Phi(b,c)$
on finite $L\times L$ square lattices up to $L=8$ for free, cylindrical,
and self-dual boundary conditions using the random-cluster tranfer matrix. 
The self-dual lattices considered
in this paper are periodic in the horizontal direction and there is
another site above the $L\times L$ square lattice, which connects
to $L$ sites on the last ($L$th) row (Figure 1).

The algorithm (random-cluster transfer matrix)
used to obtain the density of states $\Phi(b,c)$ 
is similar in spirit to that of Chen and Hu\cite{chen3}.
We consider an $L\times N$ square lattice with periodic boundary
conditions in the horizontal direction (length $L$) and free boundaries
in the vertical direction (length $N$). We define $\phi_L^{(m)}(b,c,\{t\})$
as the density of states for the $L\times m$ square lattice without 
the horizontal bonds in the $m$th row as a function of the number of
occupied bonds $b=0,1,...,2L(m-1)$, the number of clusters $c=1,2,...,Lm$,
and the top labels $\{t\}=\{t_1,t_2,...,t_L\}$ which tell whether each 
site in the $m$th row is connected to the other sites in the same row.

The first step is to calculate $\phi_L^{(2)}(b,c,\{t\})$ using the
Hoshen-Kopelman (HK) algorithm\cite{hoshen}.
The sites in the first row are labeled $1,2,...,L$ from left to right
and $L+1$ to $2L$ in the second row. 
Cluster labels $s_i$ ($i=1,2,...,2L$) are determined for each site 
and the top label $t_j$ ($j=1,2,...,L$) for the site $j+L$ in the second row
for each bond configuration.
The top label $t_j$ is the smallest number of the
set of indices $j=1,2,...,L$ for the sites in the second row belonging to 
the same cluster which includes the site $j+L$. Because $t_j\le j$, the 
maximum number of sets of top labels $\{t\}$ is $L!$. Counting the
cases $s_i=i$ gives the number of clusters $c$.

Given $\phi_L^{(m)}(b,c,\{t\})$, $\phi_L^{(m+1)}(b,c,\{t\})$ 
is calculated recursively by
\be
\phi_L^{(m+1)}(b,c,\{t\})=\sum_{\gamma=1}^{\gamma_{max}}
\sum_{b',c',\{t'\}}\phi_L^{(m)}(b',c',\{t'\})\delta(b-b'-b_g)
\delta(c-c'-\Delta c)\delta(\{t'\}\to\{t\})
\ee
for $m=2,3,...,N-1$, where $\gamma$
labels the $2^{2L}(=\gamma_{max})$ possible bond configurations in the newly
added piece, $g$, consisting of the horizontal bonds in the $m$th row
and the vertical bonds between the $m$th row and the $(m+1)$th row,
and $b_g$ is the number of occupied bonds in $g$. The sites $i$ in $g$
are labeled from left to right by $1,2,...,L$ in the $m$th row and by
$L+1,L+2,...,2L$ in the $(m+1)$th row.
We again use the HK algorithm to determine the cluster labels 
$\{s_1,s_2,...,s_{2L}\}$ and the number of clusters $c_g$ in $g$,
and the {\it updated} old top labels 
$\{\tilde{t}'\}=\{\tilde{t}'_1,\tilde{t}'_2,...,\tilde{t}'_L\}$ and
the new top labels $\{t\}=\{t_1,t_2,...,t_L\}$ making a comparison
between the cluster labels $\{s_1,s_2,...,s_L\}$ and the old top labels
$\{t'\}=\{t'_1,t'_2,...,t'_L\}$. In Eq. (12) $\Delta c$ is given by 
the Chen-Hu formula\cite{chen3}
\be
\Delta c=c_g-n-n'+n'',
\ee 
where $n$ is the number of the cluster labels satisfying $s_i=i$
for $i=1,2,...,L$, $n'$ the number of the old top labels satisfying
$t'_i=i$, and $n''$ the number of the updated old top labels satisfying
$\tilde{t}'_i=i$.

Finally, the density of states $\Phi(b,c)$ is obtained by
\be
\Phi(b,c)=\sum_{\gamma=1}^{\gamma_{max}}
\sum_{b',c',\{t'\}}\phi_L^{(N)}(b',c',\{t'\})\delta(b-b'-b_g)
\delta(c-c'-\Delta c)
\ee
with $\gamma_{max}=2^L$ and $g$ made up of the horizontal bonds
in the last ($N$th) row.

The random-cluster transfer matrix works very well, but for comparatively 
large lattices a considerable amount of memory is required to store
$\phi_L^{(N)}(b,c,\{t\})$. At the expense  of a slight increase in
the complexity of the code it is possible to reduce the memory requirements
substantially. First, the $L!$ sets of top labels include many unused sets,
such as $\{...,t_i=i,t_j=i,t_k=j,...\}$ ($i<j<k$), which account for 56.7 \%
of all sets for $L=5$ and 96.8 \% for $L=10$ and can be removed easily
from $\phi_L^{(m)}(b,c,\{t\})$. Second, we should consider the fact that
only some range of $c$ is used for a fixed $b$. For example, in 
$\phi_5^{(5)}(b,c,\{t\})$ only $c=1$ to 11 ($\delta c=11$) are needed 
for $b=24$. Here $c=1$ results from the sparsest distributions 
of 24 occupied bonds and $c=11$ from the most compact distributions. 
$\delta c\le11$ for all $b\ne24$, and $(\delta c)_{max}=11$.
We can calculate $(\delta c)_{max}$ easily for $\phi_L^{(N)}(b,c,\{t\})$
and reduce a large amount of memory. Third, $\Phi(b,c)$ can be obtained
directly from $\phi_L^{(m)}(b',c',\{t'\})$ ($m\le N-1$) with
$\gamma_{max}=2^{L+2L(N-m)}$ using Eq. (14). This method decreases
memory requirements but increases CPU time, while the former two methods 
reduce both the memory and CPU time requirements. In general, the 
random-cluster transfer matrix based on Eq. (12) is very fast, taking just 
30 seconds on a PC with one PENTIUM 100 MHz CPU to obtain $\Phi(b,c)$
on the $5\times5$ square lattice with free boundary conditions.

The density of states $\Omega_Q(E)$ is related to the density of states
$\Phi(b,c)$ by
\be
\Omega_Q(E)=\sum_{b=E}^{N_b}\sum_{c=1}^{N_s}\Phi(b,c) Q^c
{b\choose E} (-1)^{b-E}
\ee
for integer $Q$. In Eq. (15) $Q$ need not be an integer and
Eq. (15) defines the density of states $\Omega_Q(E)$ 
of the $Q$-state Potts model for noninteger $Q$.


\section{the critical Potts model}

At the ferromagnetic critical point, $a_c=1+\sqrt{Q}$,
the partition function of the $Q$-state Potts model becomes
\be
Z_{CP}=\sum_{b,c}\Phi(b,c)\biggl(\sqrt{Q}\biggr)^{b+2c},
\ee
which is a polynomial in $\sqrt{Q}$.
This defines what we refer to as the {\it critical} Potts model.
Since $b=N_s-1$, $c=1$ and $b=0$, $c=N_s$ set the lowest and highest
orders, respectively, in the polynomial, we can write Eq. (16) as
\be
Z_{CP}=w^{N_s+1}\sum_{r=0}^{N_s-1}K_r w^r,
\ee
where $w=\sqrt{Q}$. The coefficients $K_r$ of the new polynomial $Z_{CP}$ 
satisfy
\be
\sum_{r=0}^{N_s-1}K_r=2^{N_b}
\ee
and
\be
\sum_{r=0}^{N_s-1}K_r(-1)^r=0.
\ee
Table I shows the coefficients $K_r$ for the 
$8\times8$ square lattice with free boundary conditions.

In addition to the ferromagnetic critical point $a_c=1+\sqrt{Q}$,
the point $\bar{a}_c=1-\sqrt{Q}$, which is sometimes referred to the 
{\it unphysical critical point}, also maps into itself under 
the dual transformation $(\tilde{a}-1)(a-1)=Q$\cite{wu}. This leads us 
to consider the corresponding critical Potts partition function
\be
\bar{Z}_{CP}=\bar{w}^{N_s+1}\sum_{r=0}^{N_s-1}K_r \bar{w}^r,
\ee
where $\bar{w}=-w$. Evidently $\bar{Z}_{CP}$ can be obtained from
${Z}_{CP}$ simply by continuing $w$ to negative values.
With this understanding we consider $Z_{CP}(w)$ for arbitrary complex
values of $w$. Note that the map of the complex $w$ plane on to the
complex $Q$ plane is now 2-to-1.

Figure 2 shows the Potts zeros in the complex $w$ plane of the critical 
Potts model on an $8\times8$ square lattice with self-dual boundary 
conditions. The zero at $w=0$ is $N_s+1$ degenerate, and 
most of the remaining $N_s-1$ zeros lie in the half space ${\rm Re}(w)<0$. 
Several of these zeros lie on the negative real axis,
and these will map on to the positive real $Q$ axis as shown in Figure 3. 
Some of these zeros (Table II) lie at or close to the Beraha numbers
\cite{baxter3}
\be
B_n=4\cos^2{\pi\over n}
\ee
with $n=2,3,4,...$ and $0\le B_n\le4$. In the study of the phase diagram
of the Potts model Saleur\cite{saleur} assumed that 
the Potts model at the unphysical critical point, 
$\bar{a}_c=1-\sqrt{Q}$, is singular when $Q=B_n$,
and our results verify this observation.
Table II shows the Potts zeros of the critical Potts model on the
$L\times L$ square lattice which lie at or close to 
the Beraha numbers for free ($N_b=2L^2-2L$), cylindrical ($2L^2-L$), 
and self-dual ($2L^2$) boundary conditions. 
As the number of bonds, $N_b$, increases, the number of the Potts zeros 
at or close to the Beraha numbers $B_n$ increases for a fixed $L$, 
and as $L$ increases the number of the Potts zeros at or close to $B_n$
increases for any specified type of boundary conditions. 
We expect that in the thermodynamic limit the Potts zeros on 
the positive real axis cover all the Beraha numbers $B_n$ ($n=2,3,...$).

For self-dual boundary conditions there exist unexpected Potts zeros 
on the positive real axis for $Q>4$ (Table III). These zeros do not
exist for non-dual boundary conditions, and the largest of these zeros,
which we shall denote by $Q_{max}(L)$, has an interesting significance.  
Recently the partition function zeros in the 
complex temperature plane (Fisher zeros) have been studied extensively
for the Potts model\cite{kim1,maillard,martin1,martin2,martin3,wood,bhanot,alves,martin4,chen1,chen2,shrock3,creswick2,kenna2,kim3}. 
By numerical methods it has been shown
\cite{kim1,martin2,martin3,martin4,chen1,shrock3,creswick2,kim3} that
for self-dual boundary conditions the Fisher zeros of the $Q>1$ Potts models
on a finite square lattice are located on the unit circle in the complex
$p$ plane for ${\rm Re}(p)>0$, where $p=(a-1)/\sqrt{Q}$.
Chen {\it et al.}\cite{chen1} conjectured that when $Q$ reaches 
a certain critical value $\tilde{Q}_c(L)$, all Fisher zeros are located 
on the unit circle 
$|p|=1$. However, the value of $\tilde{Q}_c(L)$ and how it scales with 
$L$ were not addressed. We find that $\tilde{Q}_c(L)$ is identical to
$Q_{max}(L)$ and that $\tilde{Q}_c(L)$ increases 
with $L$ as shown in Table III. 

Figure 4 shows the Fisher zeros in the complex $p$
plane of the $Q$-state Potts model on the $4\times4$ square lattice 
with self-dual boundary conditions. For $Q=75$ the two zeros on the
negative real axis lie off the unit circle, while for $Q=76$ all the
Fisher zeros lie on the unit circle. At $Q=\tilde{Q}_c$ ($=75.37...$
for $L=4$) the two zeros lie on $p=-1$. In general, for the values of 
$Q$ (both $Q\le4$ and $Q>4$) that are determined from the Potts zeros 
on the positive real axis, two Fisher zeros always lie at $p=-1$.
$Q=1$ is exceptional in that {\it all} Fisher zeros of the one-state Potts
model lie at $p=-1$\cite{chen1}. 
Note that in Figure 4(b) the Fisher zeros are grouped
and there exists a wide gap between two neighboring groups except for
$p=-1$. Whenever all Fisher zeros lie on the unit circle, the number
of groups of zeros is $2L_x$ and the number of zeros for each group
is $L_y$, where $L_x$ and $L_y$ are the lattice sizes in the horizontal
and vertical directions, respectively.

By using the Bulirsch-Stoer (BST)
algorithm\cite{bst} we extrapolated $1/\tilde{Q}_c(L)$ 
for finite lattices to infinite size. The error estimates are twice 
the difference between the ($n-1,1$) and ($n-1,2$) approximants. 
For $\omega=1$ (the parameter of the BST algorithm) we get 
$1/\tilde{Q}_c=0.0007(8)$ and $1/\tilde{Q}_c=0.0001(7)$ 
for $\omega=2$. These results imply that in the 
thermodynamic limit all the Fisher zeros lie on the unit circle
only in the limit $Q\to\infty$\cite{chen2}. Conversely, this observation
implies that the locus of zeros in the thermodynamic limit for finite $Q$
is an open question.


\section{antiferromagnetic Potts zeros}

For antiferromagnetic interaction, $J<0$, the physical interval is
$0\le a\le 1$ ($0\le T\le\infty$). At zero temperature
($a=0$) the partition function is
\be
Z=\sum_{b,c}\Phi(b,c)(-1)^b Q^c,
\ee
which is also known as the chromatic polynomial in graph theory
\cite{shrock5,sokal4}.
Figure 5 shows the zeros of the chromatic polynomial in the complex
$Q$ plane for the $8\times8$ square 
lattice for cylindrical\cite{baxter3} and self-dual boundary conditions.
In Figure 5, except for the zeros at the Beraha numbers 0, 1 and 2
($Q=2.0000000000007$ for cylindrical boundary conditions),
the Potts zeros are distributed along curves which cut the positive real 
axis between $Q=2$ and 3. The intersection of the locus of the Potts zeros
with the real axis depends on the boundary condition: for $L=8$  and
cylindrical boundary conditions we have $Q=2.551073$, while for self-dual
boundary conditions we find a pair of zeros at $Q=2.636589$ and 2.645969 
which are slightly larger than the fifth Beraha number $B_5=2.618034$.
For the $7\times7$ self-dual lattice these zeros lie at $Q=2.621577$
and $Q=2.684634$ (Figure 6). In addition for $L=7$ there are isolated
zeros on the real axis at the Beraha numbers $B_2=0$, $B_3=1$, and 
$B_4=2$, and an additional zero appears at $B_6=3$ (Figure 6). 
$Q=3$ corresponds to the critical value $Q_c$\cite{shrock4,biggs} which 
separates the region ($Q\le3$) with  antiferromagnetically ordered 
ground-states from the region ($Q>3$) of disordered states at $T=0$.
Here we generalize this concept to finite temperatures and define
$Q_c(a)$ to be the value of $Q$ for a given value of $a$ below which
there is antiferromagnetic order.  
Because four colors are needed to color an $L\times L$ square lattice with 
self-dual boundary conditions such that no two nearest neighbors have 
the same color, there exists a trivial Potts zero 
at $Q_c=3$ when $L=3,5,7,...$ .

Figure 6 shows the Potts zeros of the dichromatic polynomial at several 
temperatures for the $7\times7$ lattice with self-dual boundary conditions.
As $a$ is increased the zeros move toward the origin and 
converge on the point $Q=0$ for $a=1$\cite{shrock5}.
The antiferromagnetic critical point is given by
$a_c(Q)=\sqrt{4-Q}-1$\cite{baxter2,kim1}, from which we have
\be
Q_c(a)=(1-a)(a+3).
\ee
Table IV shows the Potts zeros $Q_c(L)$ on ($L=4,6,8$) or closest 
to ($L=3,5,7$)
the positive real axis for $a=0.5$. From the BST extrapolation we 
obtained $Q_c=1.78(18)$ (from $L=4,6,8$) and $Q_c=1.77(36)-0.01(3)i$
(from $L=3,5,7$) in agreement with Eq. (23).
Figure 7 compares Eq. (23) (continuous curve) with the BST estimates
from $Q_c(a,L)$ for $L=3,5,7$ and self-dual boundary conditions
for several values of $a$.  


\section{ferromagnetic Potts zeros}

For ferromagnetic interaction, $J>0$, the physical interval is 
$a=[1,\infty]$ ($T=[\infty,0]$). Figure 8 shows the Potts zeros 
of the dichromatic polynomial on $L\times L$ lattices with cylindrical
boundary conditions for $a=1+\sqrt{2}=2.414...$ and 
$a=1+\sqrt{3}=2.732...$ . For free and self-dual boundary conditions
the distribution of the Potts zeros is similar to that for cylindrical
boundary conditions. Unlike the antiferromagnetic Potts zeros which
are distributed mainly in the ${\rm Re}(Q)>0$ region (Figures 5 and 6),
many ferromagnetic Potts zeros lie in the ${\rm Re}(Q)<0$ region.
With the exception of the trivial zero at $Q=0$ the ferromagnetic Potts 
zeros are distributed along a single curve which moves away from the origin 
as $a$ increases. There is no zero on the positive real axis, but the zero 
$Q_1(a,L)$ closest to the positive real axis approaches the real axis 
as $L$ increases. As in the Yang-Lee theory\cite{yang}, we expect 
$Q_1(a,L)\to Q_c(a)$ in the limit $L\to\infty$.
Table V shows the BST estimates from $Q_1(a,L)$ at $a=1+\sqrt{2}$ and 
$1+\sqrt{3}$ for different boundary conditions, suggesting that the locus
of the Potts zeros cuts the positive real axis at $Q_c=2$ and 3, 
respectively, in the thermodynamic limit. From the ferromagnetic critical 
point, $a_c(Q)=1+\sqrt{Q}$, we obtain
\be
Q_c(a)=(a-1)^2,
\ee
which we have confirmed for $a=1+\sqrt{2}$ and $1+\sqrt{3}$ and other
values of $a>1$ (Figure 9).

The behavior of the closest zero $Q_1(a,L)$ suggests a new scaling exponent
$y_q$ defined as
\be
Q_1(a,L)\simeq Q_c(a)+A L^{-y_q}.
\ee
For finite lattices we define\cite{kim1,bhanot,alves,kim3,creswick3}
\be
y_q(L)=-{{\rm ln \{Im}[Q_1(L+1)]/{\rm Im}[Q_1(L)]\}\over{\rm ln}[(L+1)/L]}.
\ee
The exponent $y_q$ is to the Potts zeros in the complex $Q$ plane what
the thermal exponent $y_t$ (or the magnetic exponent $y_h$)
is to the Fisher zeros in the complex temperature plane
(the Yang-Lee zeros in the complex magnetic-field plane). 
Figure 10 shows the BST estimates from $y_q(L)$ for $a=2$ ($Q_c=1$),
$1+\sqrt{2}$ ($Q_c=2$), $1+\sqrt{3}$ ($Q_c=3$), and 3 ($Q_c=4$).
The exponent $y_q$ increases as $a$ (or $Q_c$) increases.
Figure 10 compares our results for $y_q$ versus $a_c(Q)$ with the den Nijs
formula\cite{nienhuis,kim1} for the thermal exponent $y_t(a_c(Q))$
of the ferromagnetic Potts model.
Clearly the general behaviors of $y_q$ and $y_t$ with $a_c(Q)$ are similar;
these initial results are of insufficient precision to settle the
question that $y_q=y_t$ or not. 


\section{Fisher zeros and Potts antiferromagnets}

For antiferromagnetic interaction $J<0$ the physical interval is
$0\le a=e^{\beta J}\le1$ ($0\le T\le\infty$), which corresponds to 
\be
{-1\over\sqrt{Q}}\le p={a-1\over\sqrt{Q}}\le0.
\ee
From the exact partition functions, Eqs. (4) and (11), we have evaluated
Fisher zeros of the Potts model. Figure 11 shows the Fisher zeros in the
complex $p$ plane of the three-state Potts model on a $12\times12$ square
lattice with free boundary conditions. The Fisher zeros in the complex
$p$ plane of the $Q$-state Potts model for several values of non-integer
$Q$ have been shown for the $8\times8$ square lattice with self-dual
boundary conditions\cite{kim1}. Figure 12 shows the Fisher
zeros in the complex $p$ plane of the $Q=2.5$ Potts model on 
an $8\times8$ square lattice with free boundary conditions. 
In Figures 11 and 12 there is a group of complex zeros approaching 
the antiferromagnetic critical point $a_c=\sqrt{4-Q}-1$, equivalently,
$p_c=(a_c-1)/\sqrt{Q}$ and crossing the real axis at this critical point
in the thermodynamic limit\cite{kim1}. For an $L\times L$ square lattice
$a_c(L)$ or $p_c(L)$ denotes the closest zero to the antiferromagnetic
critical point or edge singularity.
Based on the finite-size scaling law of the partition function zeros
near the critical point\cite{itzykson,glasser} we expect
\be
{\rm Im}[a_c(L)]\sim L^{-y_t},
\ee
from which we can estimate the thermal exponent $y_t(L)$ for finite 
lattices as\cite{kim1,bhanot,alves,kim3,creswick3}
\be
y_t(L)=-{\ln \{{\rm Im}[a_c(L+1)]/{\rm Im}[a_c(L)]\}\over\ln[(L+1)/L]}.
\ee
Table VI shows the thermal exponents $y_t(L)$ of the Ising ($Q=2$)
antiferromagnet and the three-state Potts antiferromagnet
for free boundary conditions.
By using the BST algorithm
we extrapolated our results for $y_t(L)$ to infinite size
for $2\le Q\le3$.
Figure 13 shows the thermal exponent $y_t$ of the Potts antiferromagnet
by the BST estimates with $\omega=1$ (the parameter of the BST algorithm)
for free boundary conditions. For the BST extrapolation of finite-size
results of the Potts antiferromagnet we prefer free boundary condition
to other boundary conditions. The reason for this is that even though
finite size effects are larger for free than cylindrical boundary 
conditions, the edge singularity approaches the critical point 
monotonically only if we consider a sequence of lattices with $L$ even.
For free boundary conditions this is not a problem and the increased
effectiveness of the BST algorithm with longer sequences more than 
compensates the stronger finite-size effects\cite{shrock1,kim1}.
In Figure 13 there are two BST estimates for $Q=3$. 
The upper estimate resulted from data
for $L=3\sim8$, while the lower one uses $L=3\sim12$. In Figure 13
the continuous curve is the fit to the BST estimates with  
\be
y_t={1+Au+Bu^2\over C+Du},
\ee
where 
\be
u=-{2\over\pi}\cos^{-1}{\sqrt{Q}\over2},
\ee
and $A=-2.2821$, $B=-7.4390$, $C=3.9818$, and $D=7.4011$.
The variable $u$ arises naturally in the expressions for the free energy
$f_c(-{\pi\over2}u)$ at the ferromagnetic\cite{baxter1} and 
antiferromagnetic\cite{baxter2} critical temperatures, and in the critical
exponents $y_t$\cite{nienhuis,blote,kim1} and $y_h$\cite{pearson,blote}
of the ferromagnetic Potts model.
The form used in Eq. (30) has also been used to describe the critical 
exponent $y_h$ of the ferromagnetic Potts model\cite{pearson}. 

The BST estimates of the thermal exponent $y_t$ for $Q<3$
are insensitive to the parameter of the BST algorithm, $\omega$.
However, as $Q$ approaches 3 
the BST results for the three-state Potts antiferromagnet are very
sensitive to $\omega$. For example, we obtained $y_t=0.50(8)$ for $\omega=1$, 
$y_t=0.59(2)$ for $\omega=2$, and $y_t=0.60(2)$ for $\omega=3$
using data for $L=3\sim12$. The BST estimates of the thermal exponents
of the $Q$-state Potts antiferromagnets for non-integer $Q$ are also
sensitive to $\omega$ when $Q\approx3$.  Recently Ferreira and Sokal
\cite{sokal1,sokal3} have suggested the correlation
length for the three-state Potts antiferromagnet has the form
\be
\xi\sim a^{-1/y_t}(-\ln a)^r(1+c_1 a+c_2 a^2+...)
\ee
with $y_t={1\over2}$\cite{saleur,sokal1,sokal3}, $r\approx1$, and 
$c_1\approx15$.
For $Q=3$ the sensitivity of the BST estimates of the thermal exponent
to $\omega$ may result from this kind of logarithmic behavior.  

Figure 14 shows the BST results extrapolated from ${\rm Im}[a_c(L)]$ for
$L=3\sim12$ of the three-state Potts antiferromagnet with free boundary
conditions as a function of $\omega$ along with the error estimates.
When we use the BST algorithm to estimate a critical point,
the best value of the free parameter $\omega$ is the critical 
exponent $y_t$\cite{bst}. We have obtained the desired result 
${\rm Im}(a_c)=0$ for $\omega=0.5$ which strongly suggests $y_t=0.5$.  


\section{conclusion}

We have introduced the random-cluster transfer matrix 
to calculate exact integer values for the density of states $\Phi(b,c)$,
from which the exact partition function $Z(a,Q)$ can be obtained 
for any value of $Q$, even for complex $Q$. 
We have found a subset of the zeros of the partition function of the
critical Potts model in the complex $w=\sqrt{Q}$ plane which lie close
to or at the Beraha numbers on the negative real axis. The largest of
these determines $\tilde{Q}_c(L)$, the value of $Q$ above which the locus
of zeros in the complex $p$ plane lie on the unit circle. By studying
the scaling behavior of $\tilde{Q}_c(L)$ with $L$ we find that 
$1/\tilde{Q}_c(L)\to0$ as $L\to\infty$, indicating that all the zeros
do not lie strictly on the unit circle in the thermodynamic limit.

We have studied the locus of zeros of the dichromatic polynomials in
both the ferromagnetic and antiferromagnetic cases and find that the 
Yang-Lee mechanism is at work in the complex $Q$ plane. We find
$Q_c^{AF}(a)=(1-a)(a+3)$ in agreement with Baxter\cite{baxter2,kim1},
and $Q_c^{FM}(a)=(a-1)^2$ which is well known from duality arguments.
Finally, we introduce a new finite-size scaling exponent, $y_q$,
which describes the approach of the edge singularity in the complex
$Q$ plane to the critical point as $L\to\infty$. We find that $y_q$ 
varies with $Q$ in much the same way as the thermal exponent $y_t$ 
of the ferromagnetic Potts model, but as yet we have not
established a functional relation between $y_t$ and $y_q$.

We have also described the microcanonical transfer matrix to
evaluate exact integer values for the density of states $\Omega_Q(E)$
for the $Q$-state Potts model.
From the densities of states $\Phi(b,c)$ and $\Omega_Q(E)$
the partition functions $Z(a,Q)$ and $Z_Q(a)$ 
are obtained at any temperature $a$. Using the Fisher zeros of the exact
partition functions we have estimated the thermal exponents $y_t$
of the square-lattice $Q$-state Potts antiferromagnets for $2\le Q\le3$. 
For $Q<3$ the BST estimates are quite stable and $y_t$ is well
approximated by a simple algebraic function of  
$u=-{2\over\pi}\cos^{-1}{\sqrt{Q}\over2}$. However, as $Q$ approaches 3,
the BST estimates become sensitive to the choice of the scaling exponent
$\omega$ and to the data set used. Logarithmic or other corrections to
scaling may be responsible for this behavior. For $3\le L\le8$ and using
the fit from data for $Q<3$ we estimate $y_t(Q=3)\simeq0.60(2)$,
whereas if we include calculations for $L$ up to 12 we find 
$y_t(3)\simeq0.50(8)$, in agreement with the leading scaling behavior
suggested by Ferreira and Sokal\cite{sokal1,sokal3}. We hope to resolve
this issue by extending our exact calculations to larger lattices 
both exactly and by evaluating the density of states by microcanonical
Monte Carlo sampling\cite{lee}.


\begin{center}
ACKNOWLEDGMENTS
\end{center}
We thank Profs. Chin-Kun Hu and F. Y. Wu for their warm hospitality
during our stay in the Institute of Physics of the Academia Sinica,
where part of this work was carried out.
We are grateful to Prof. Robert Shrock for valuable discussions and 
for making available his two preprints\cite{shrock2,shrock5} 
before publication.
S.-Y. K. thanks Drs. Chi-Ning Chen, Jau-Ann Chen, and Youngho Park
and Prof. Nickolay Sh. Izmailian for their kind hospitality extended
to him at the Institute of Physics of the Academia Sinica. 



\begin{table}
\caption{The coefficients $K_r$ of the partition function $Z_{CP}$ of
the critical Potts model on the $8\times8$ square lattice with free
boundary conditions.}
\begin{tabular}{crcr}
$r$ &$K_r$\ \ \ \ \ \ \ \ \ \ \ \ \ &$r$ &$K_r$\ \ \ \ \ \ \ \ \ \ \ \ \  \\
\hline
0  &126231322912498539682594816       &1  &2561398756299931321297272832 \\ 
2  &25524986518920425393717379072     &3  &166557700763955734137534296320 \\
4  &800610370286991686735405550336    &5  &3023834586769553668673015126432 \\
6  &9347575153984981720573769774608   &7  &24326213916516119921387986971009 \\
8  &54404758441262921869365590686720  &9  &106224421227588059984113069365972 \\
10 &183329627865230663968273103188608 &11 &282506930412461406319413706064154 \\
12 &391942582489345467968147273830784 &13 &492998772987796894034162031881014 \\
14 &565568818070192070648821897874128 &15 &594803437106450324737629079389339 \\
16 &576045479726330572980576680006144 &17 &515761419835859402146512922316166 \\
18 &428419763789360447590812451240080 &19 &331188758886170694649818860535541 \\
20 &238937966305748243499621822108592 &21 &161285868900631598864845612258887 \\
22 &102094428513780610351844031072160 &23 &60729794216206721605782144017468 \\
24 &34010305186209829834846747925664  &25 &17962439609348242109957007244868 \\
26 &8960463658391600957849394069728   &27 &4227668735828771561070342983222 \\
28 &1888880629020154547292686697440   &29 &800023985396669919928624375932 \\
30 &321508677911960109772525527808    &31 &122688547769932427716252035294 \\
32 &44483696316227122956909056000     &33 &15331317278052765348109117036 \\
34 &5024202380355112158475486704      &35 &1565743527537870861554921235 \\
36 &464007025651505425890675200       &37 &130734234800779492211596986 \\
38 &35006515754308767635423136        &39 &8903442105259073008726006 \\
40 &2149257909558929021370016         &41 &491955405372613275069456 \\
42 &106650313357232985654928          &43 &21867081986237184782295 \\
44 &4233470330438712180496            &45 &772403311175092063841 \\
46 &132514803950430984480             &47 &21322374026497257618 \\
48 &3208188678305076656               &49 &449814829279725547 \\
50 &58534057491001584                 &51 &7036231117685951 \\
52 &776998275543312                   &53 &78304124284593 \\
54 &7144741728032                     &55 &584538167122 \\
56 &42365906128                       &57 &2678567507 \\
58 &144763280                         &59 &6504139 \\
60 &233296                            &61 &6265 \\
62 &112                               &63 &1 \\
\end{tabular}
\end{table}

\begin{table}
\caption{The Potts zeros on the positive real $Q$ axis of the critical 
Potts model which lie at or close to the Beraha numbers 
$B_n$ ($n=2,3,...$).}
\begin{tabular}{lcccc}
boundary condition &free &cylindrical &self-dual &self-dual \\
\hline
system size &$8\times8$ &$8\times8$ &$5\times5$ &$8\times8$ \\
\hline
$B_2=0$ &0 &0 &0 &0 \\
$B_3=1$ &1 &1 &1 &1 \\ 
$B_4=2$ &2.000000 &2.000000 &2.000000 &2.000000 \\
$B_5=2.618034$  &2.618034 &2.618034 &2.618055 &2.618034 \\
$B_6=3$         &3.000031 &3.000000 &2.992072 &3.000000 \\
$B_7=3.246980$  &3.226656 &3.246976 &         &3.246980 \\
$B_8=3.414214$  &         &3.415672 &3.412158 &3.414685 \\
$B_9=3.532089$  &         &3.521330 &         &3.524855 \\
$B_{10}=3.618034$ &         &         &         &3.618701 \\
$B_{16}=3.847759$ &         &         &         &3.839893 \\
$B_{30}=3.956295$ &         &         &3.957208 &         \\  
$B_{64}=3.990369$ &         &         &         &3.990438 \\
\end{tabular}
\end{table}

\begin{table}
\caption{The Potts zeros on the positive real axis for $Q>4$ for the
$L\times L$ square lattice with self-dual boundary conditions.}
\begin{tabular}{rrrrr}
$L=4$\ \ \ \ &5\ \ \ \ \ \ &6\ \ \ \ \ \ &7\ \ \ \ \ \ &8\ \ \ \ \ \ \\
\hline
75.373518 &185.886317 &395.130118 &754.036414 &1324.684018 \\
 7.566911 &            &21.911010  &40.294754   &66.309209 \\
          &            &            &6.401881   &15.678097 \\ 
          &            &            &            &5.326082 \\
\end{tabular}
\end{table}

\begin{table}
\caption{The Potts zeros on or closest to the positive real axis
for the $L\times L$ square lattice with self-dual boundary 
conditions at $a=0.5$.}
\begin{tabular}{cccc}
$L$ &$Q_c(L)$ &$L$ &$Q_c(L)$ \\
\hline
3 &$1.279400+0.161071i$  &4 &1.441800 \\
5 &$1.499871+0.0695198i$ &6 &1.574011 \\
7 &$1.583953+0.0407605i$ &8 &1.632666 \\
\end{tabular}
\end{table}

\begin{table}
\caption{The BST estimates from $Q_1(a,L)$ for different boundary 
conditions.}
\begin{tabular}{cccc}
$a$ &free &cylindrical &self-dual \\
\hline
$1+\sqrt{2}$ &$1.90(10)+0.09(24)i$ &$1.94(7)+0.15(14)i$ &$1.95(10)+0.13(11)i$ \\
$1+\sqrt{3}$ &$2.84(9)-0.13(31)i$  &$2.88(2)+0.00(15)i$ &$2.89(8)-0.03(10)i$  \\ 
\end{tabular}
\end{table}

\begin{table}
\caption{The thermal exponents $y_t(L)$ of the $Q$-state Potts 
antiferromagnets for $Q=2$ and $Q=3$ with free boundary 
conditions. The last row is the BST extrapolation
with $\omega=1$ to infinite size.}
\begin{tabular}{ccc}
$L$ &$y_t(L)$ ($Q=2$) &$y_t(L)$ ($Q=3$) \\
\hline
3  &0.859670530424 &0.672417300113 \\ 
4  &0.882900616441 &0.840771366429 \\
5  &0.895500892567 &0.750192805568 \\
6  &0.904846051999 &0.714132507277 \\
7  &0.912493138251 &0.694522575800 \\ 
8  &0.918981910221 &0.681414203729 \\
9  &0.924586147759 &0.671514256321 \\
10 &0.929481322004 &0.663473505003 \\
11 &0.933794047470 &0.656641075731 \\
$\infty$ &1.000005(9) &0.50(8) \\
\end{tabular}
\end{table}    
        

\begin{figure}
\epsfbox{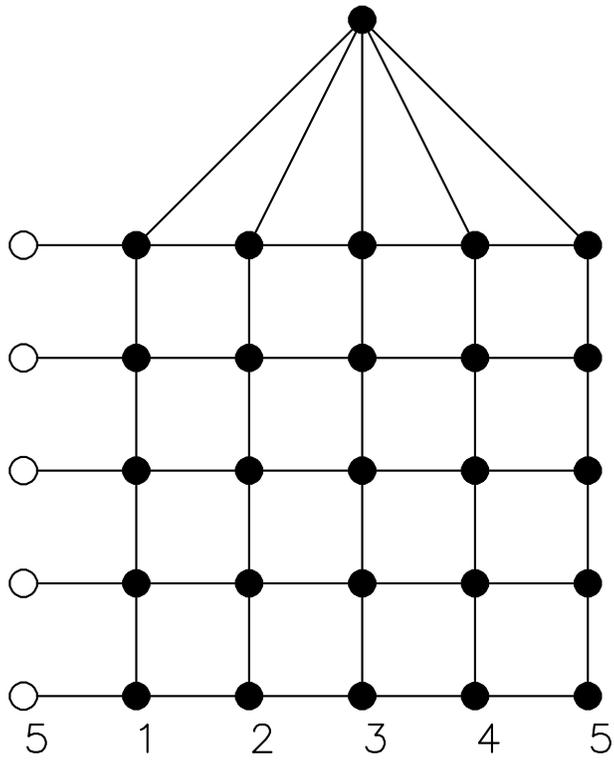}
\caption{$5\times5$ square lattice with self-dual boundary conditions.}
\end{figure}

\begin{figure}
\epsfbox{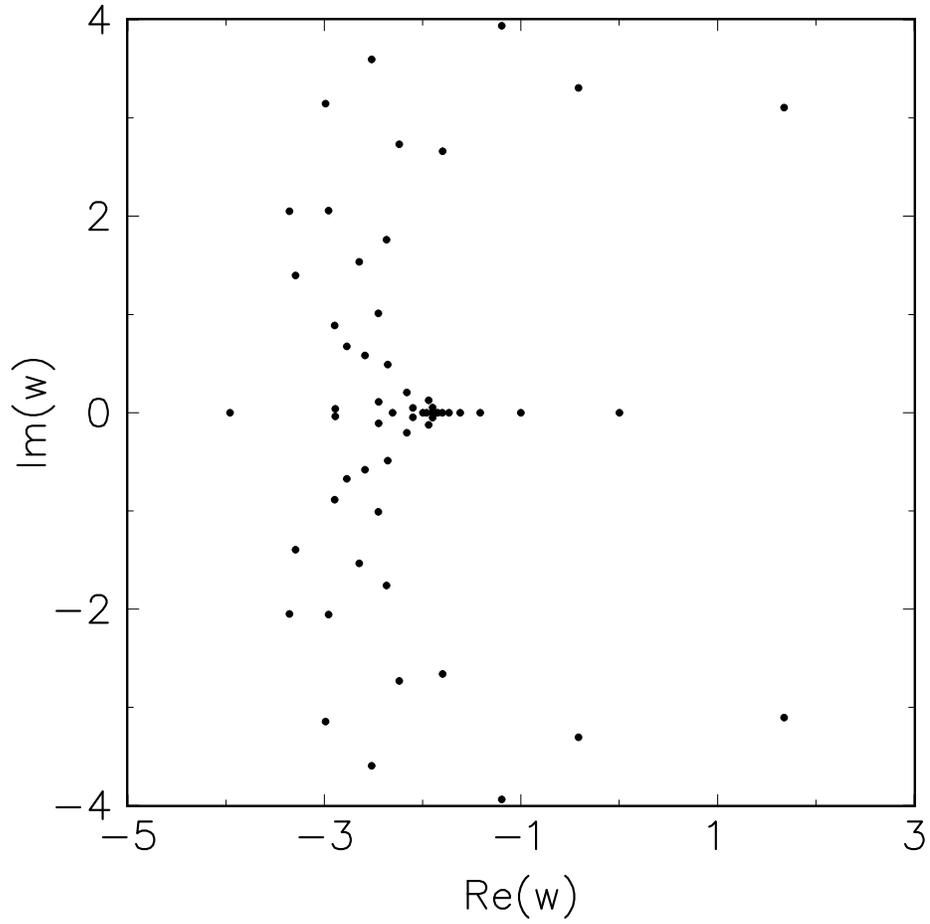}
\caption{Potts zeros in the complex $w$ ($=Q^{1\over2}$) plane of the 
partition function $Z_{CP}$ for the $8\times8$ square lattice with self-dual 
boundary conditions.}
\end{figure}

\begin{figure}
\epsfbox{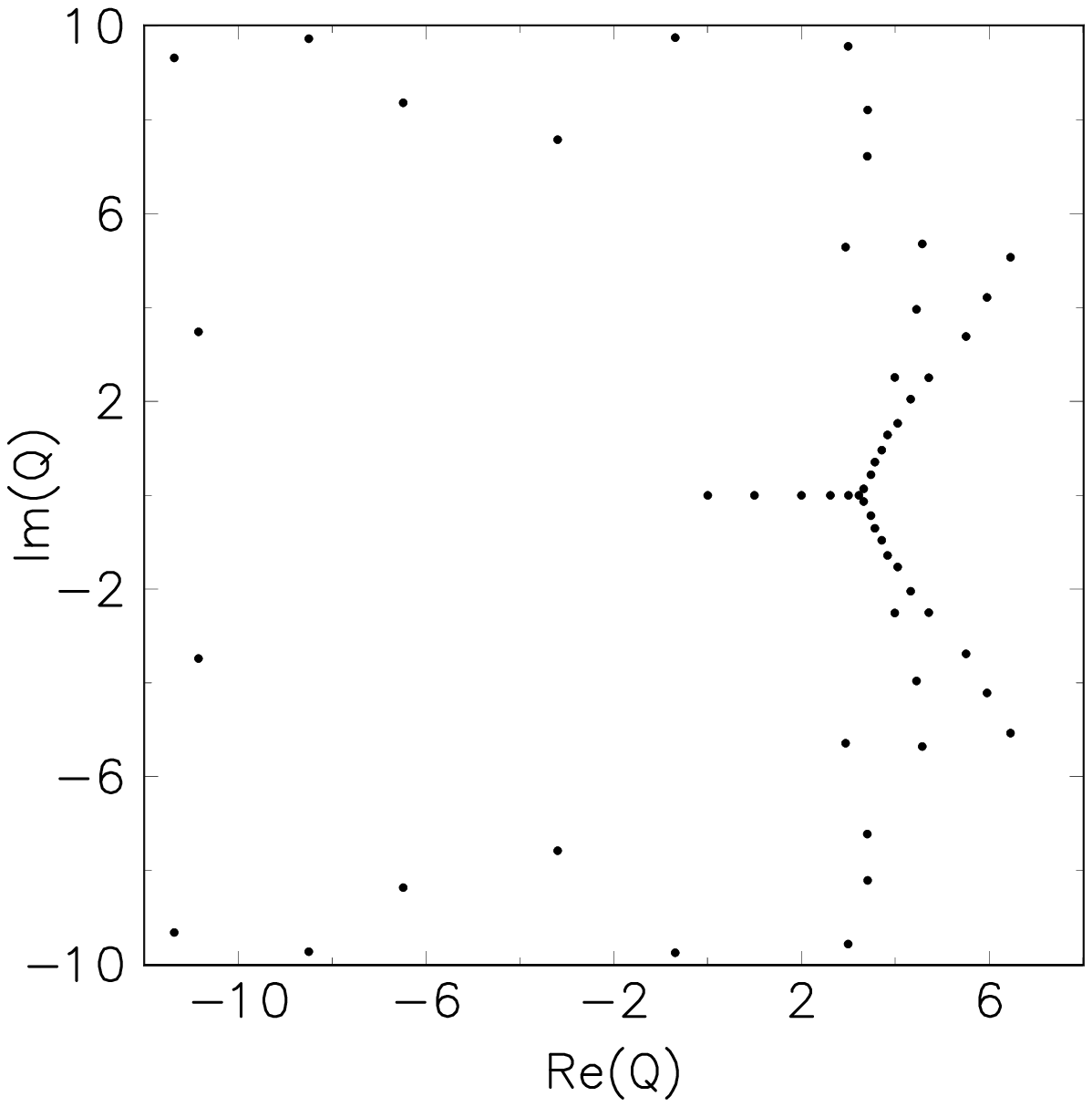}
\caption{Potts zeros in the complex $Q$ plane of the critical Potts model
for the $8\times8$ square lattice with free boundary conditions.}
\end{figure}

\begin{figure}
\epsfbox{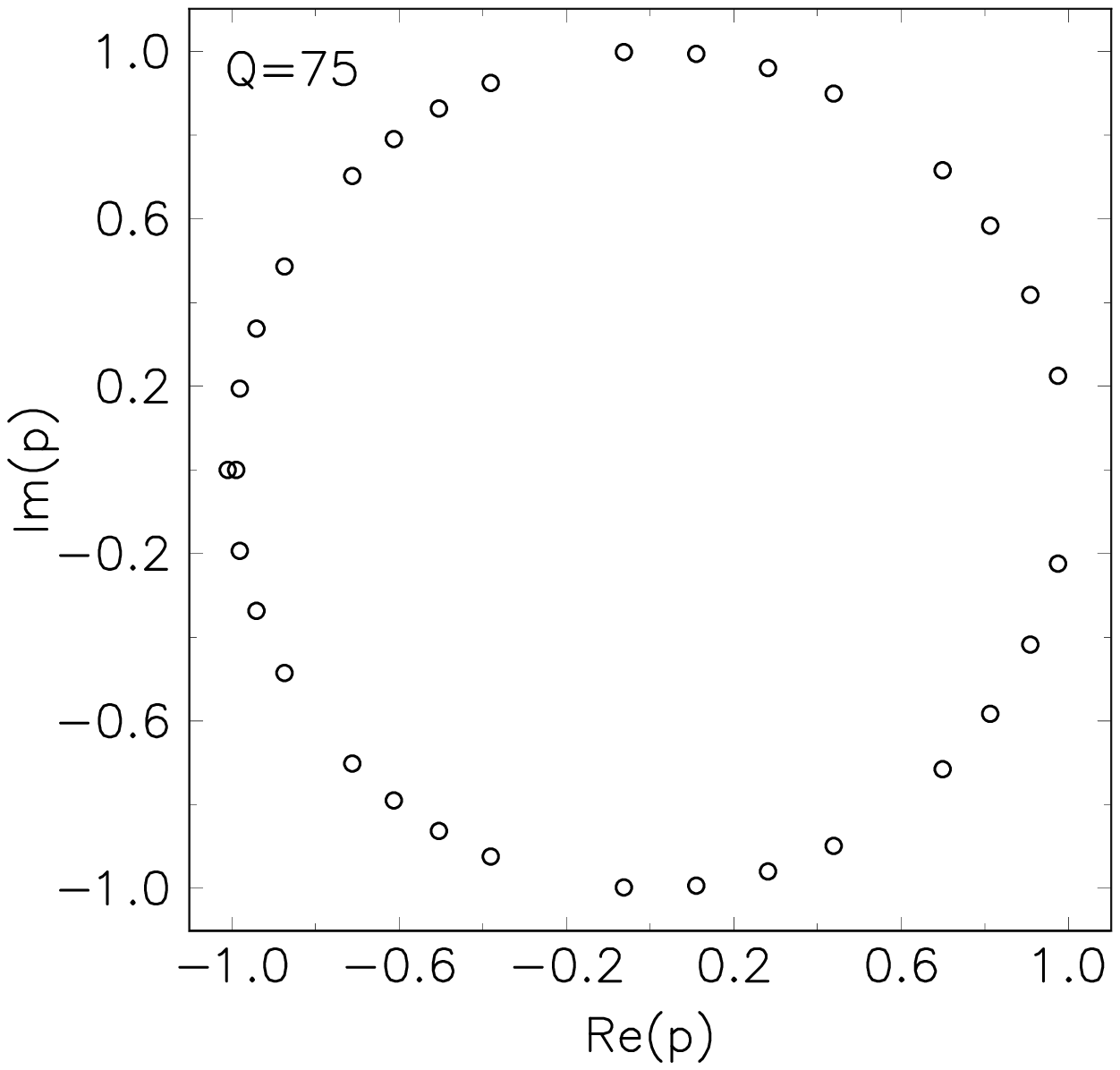}
\epsfbox{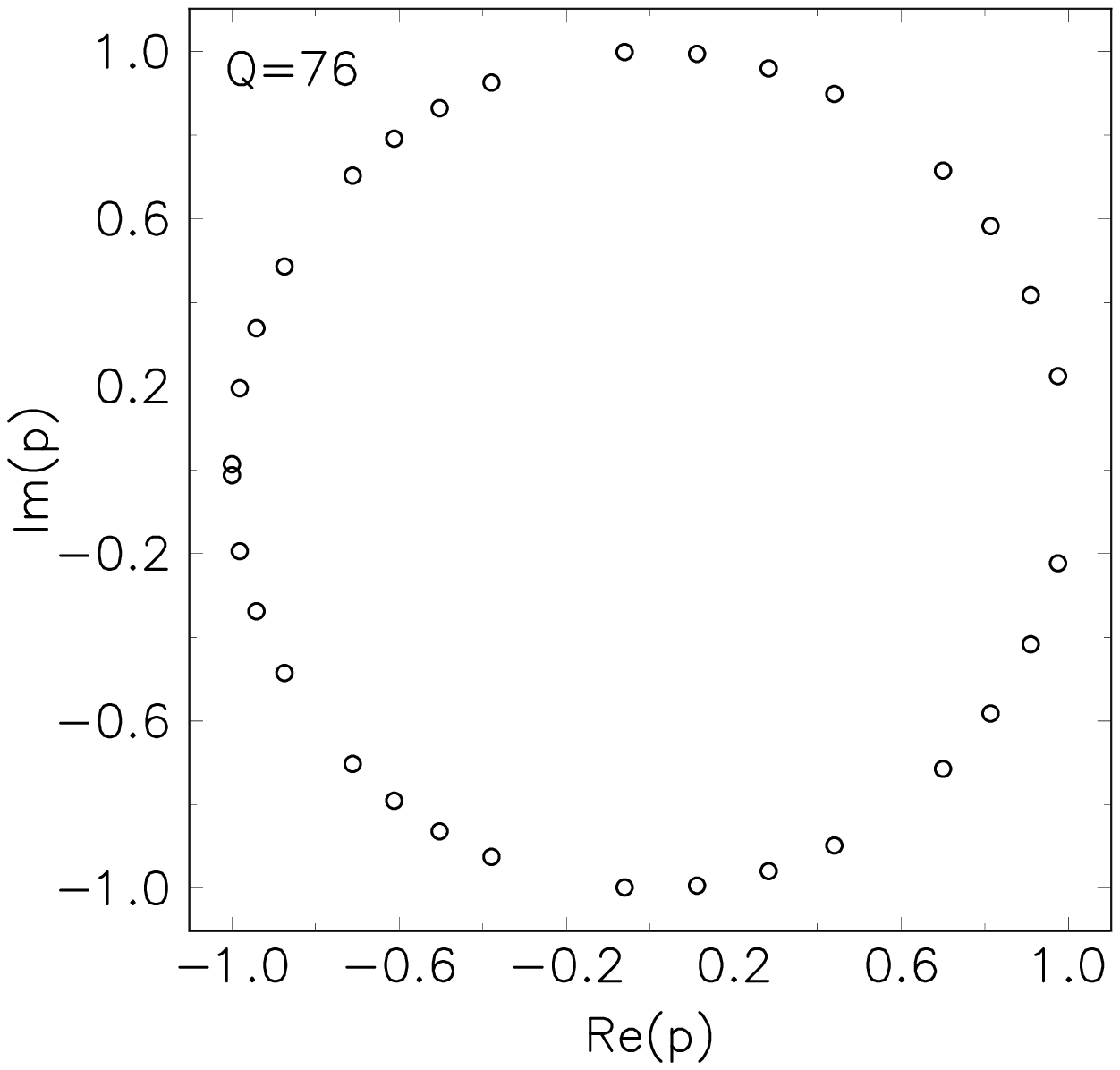}
\caption{Fisher zeros in the complex $p$ plane of the $Q$-state Potts
model on the $4\times4$ square lattice with self-dual boundary 
conditions for (a) $Q=75$ and (b) $Q=76$.}
\end{figure}

\begin{figure}
\epsfbox{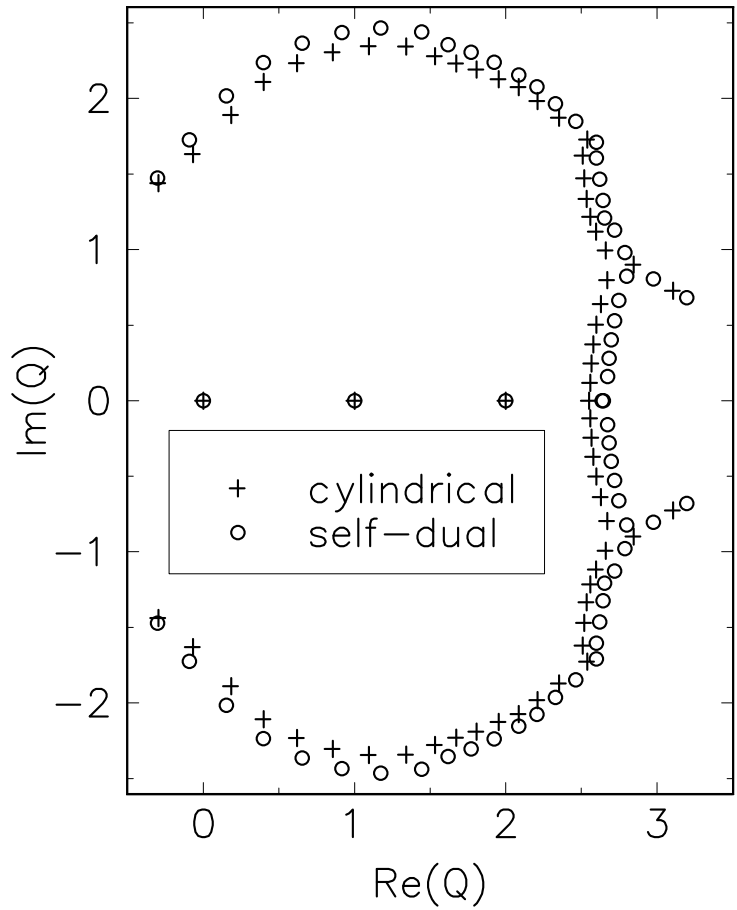}
\caption{Potts zeros in the complex $Q$ plane of the chromatic polynomial
on the $8\times8$ square lattice for cylindrical [15] and 
self-dual boundary conditions.}
\end{figure}

\begin{figure}
\epsfbox{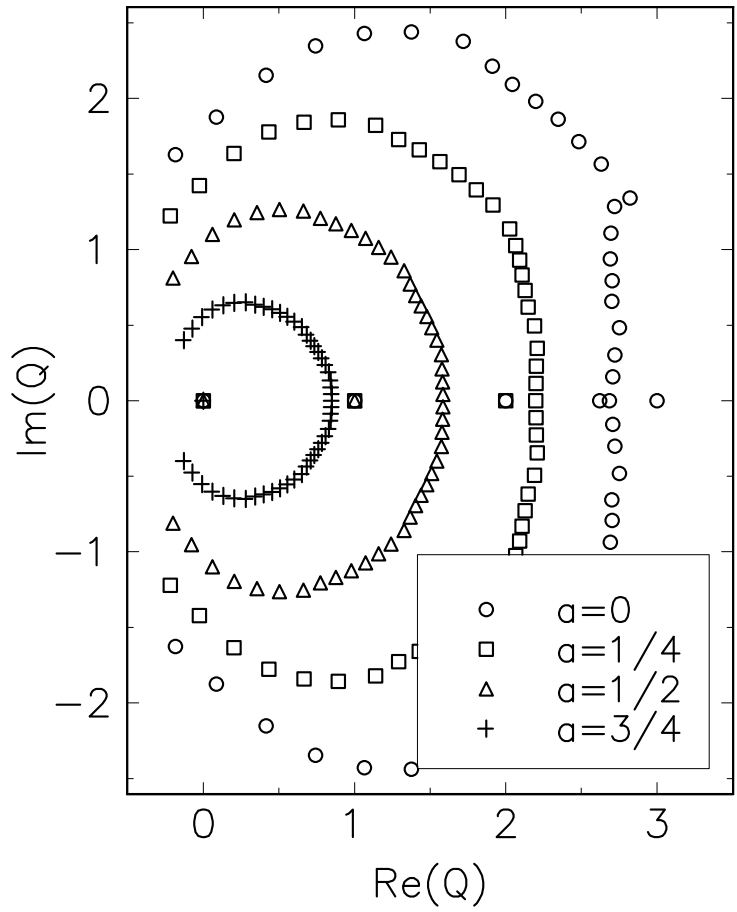}
\caption{Potts zeros of the dichromatic polynomial for $0\le a\le1$
on the $7\times7$ square lattice with self-dual boundary conditions.}
\end{figure}

\begin{figure}
\epsfbox{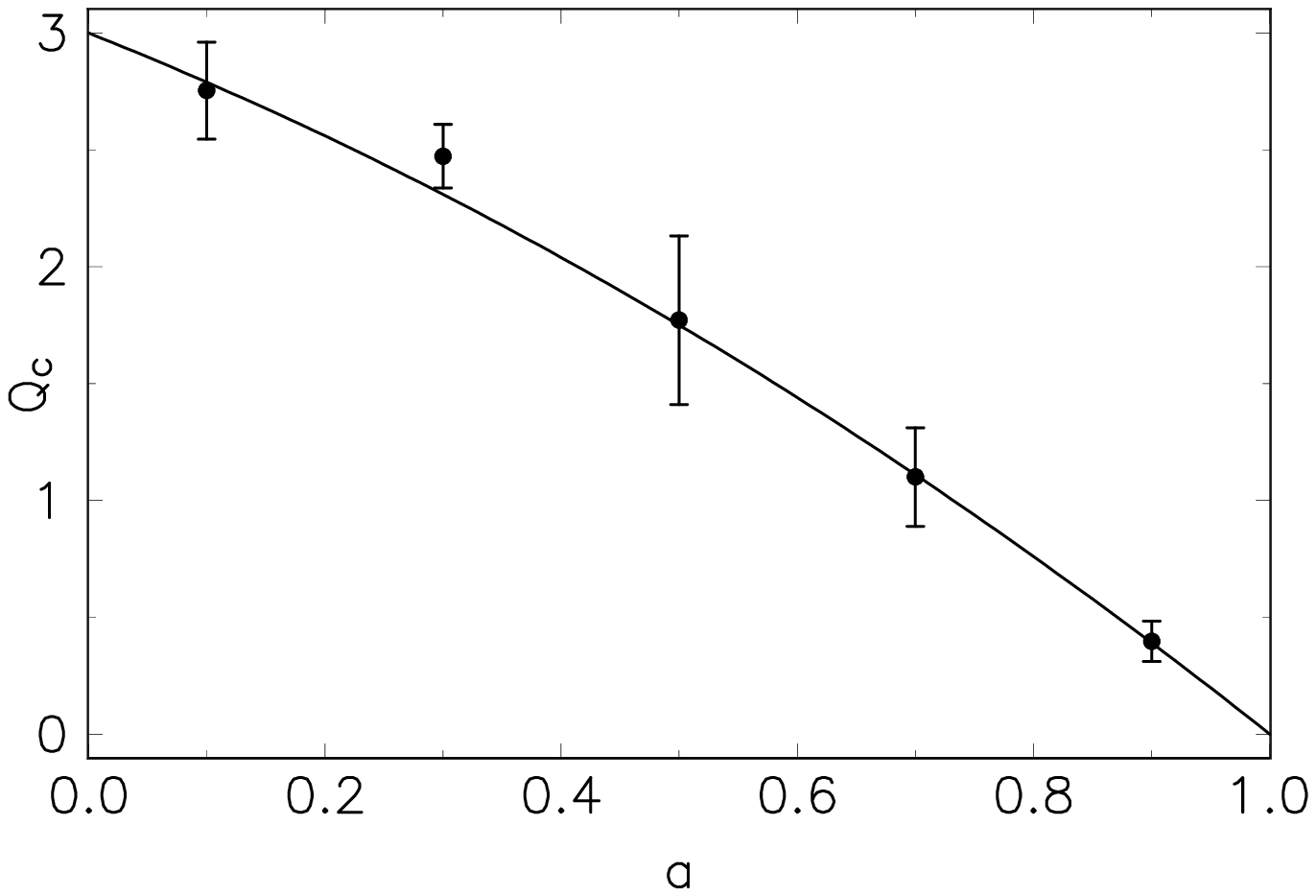}
\caption{BST extrapolation of $Q_c^{AF}$ as a function of $a$ for self-dual
boundary conditions. The continuous curve is given by $Q_c=(1-a)(a+3)$.}
\end{figure}

\begin{figure}
\epsfbox{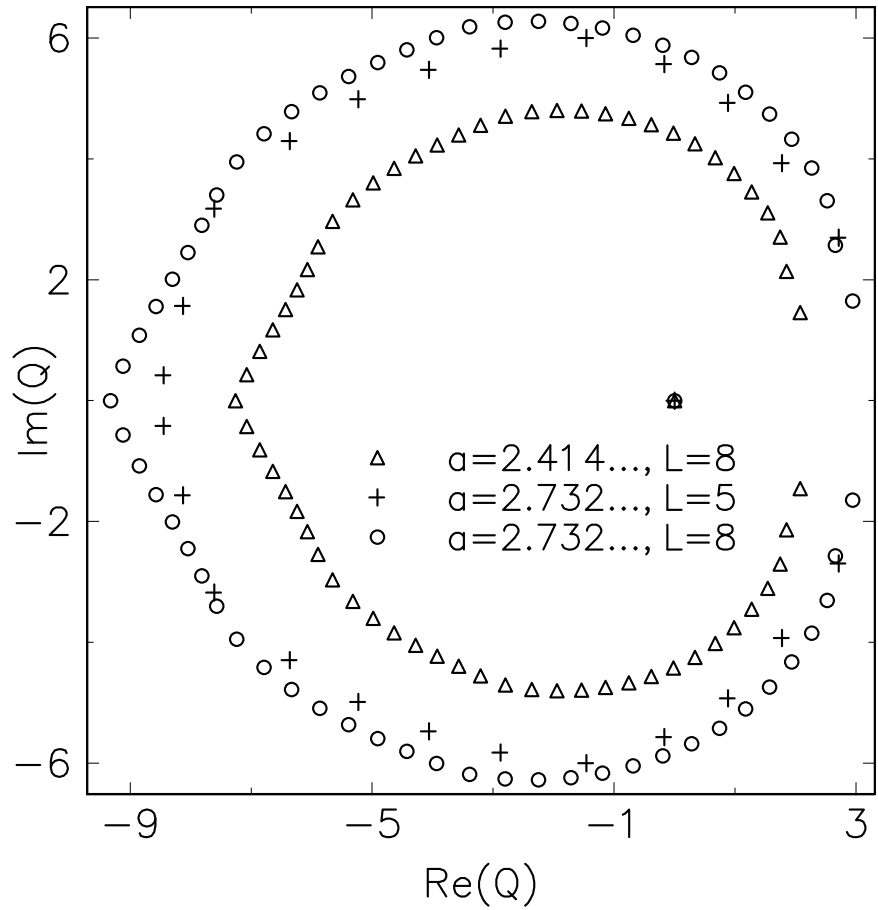}
\caption{Potts zeros in the complex $Q$ plane of the dichromatic 
polynomial on the $L\times L$ square lattices with cylindrical boundary 
conditions for $a>1$.}
\end{figure}

\begin{figure}
\epsfbox{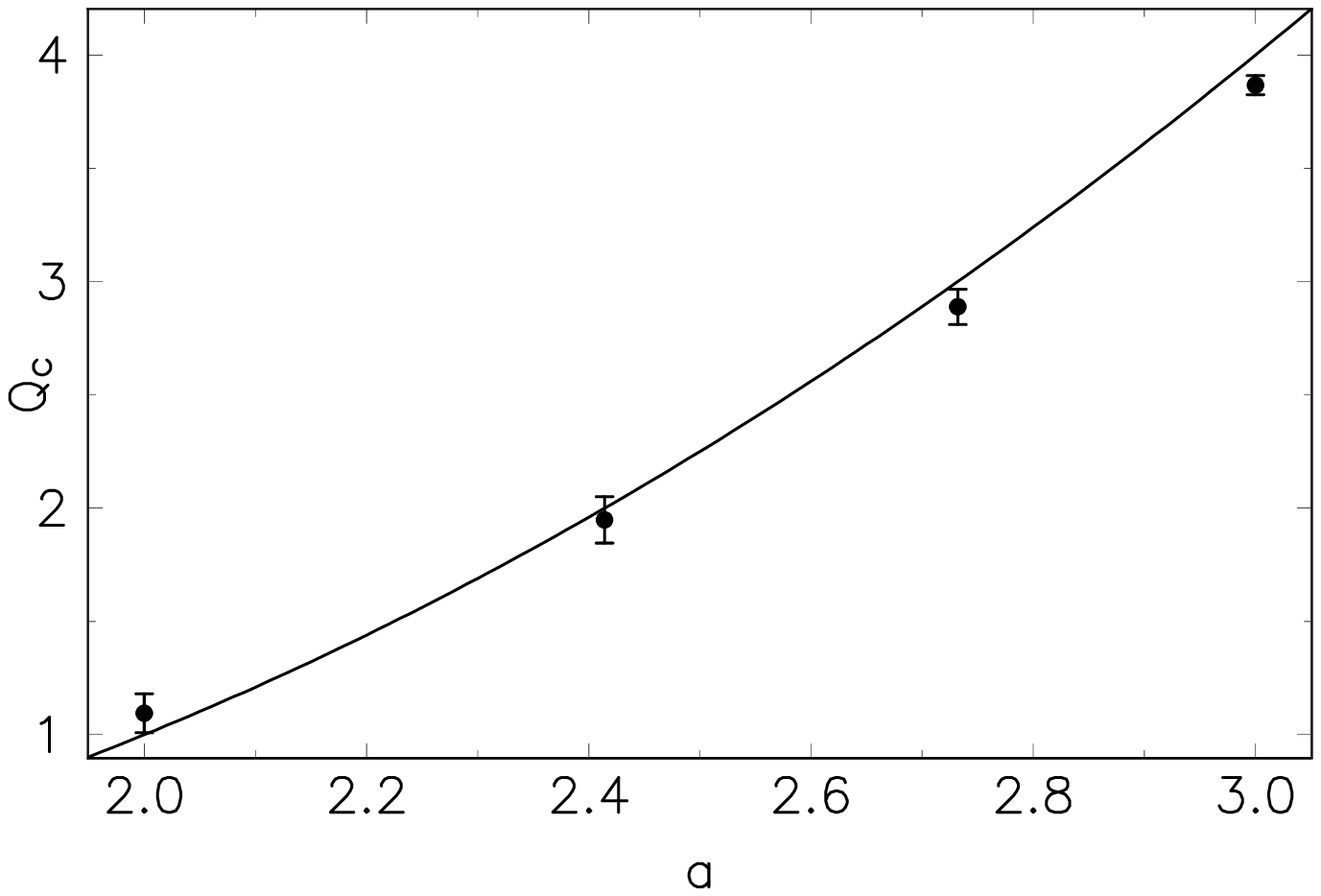}
\caption{BST extrapolation of $Q_c^{FM}$ as a function of $a$ for self-dual
boundary conditions. The continuous curve is given by $Q_c=(a-1)^2$.}
\end{figure}

\begin{figure}
\epsfbox{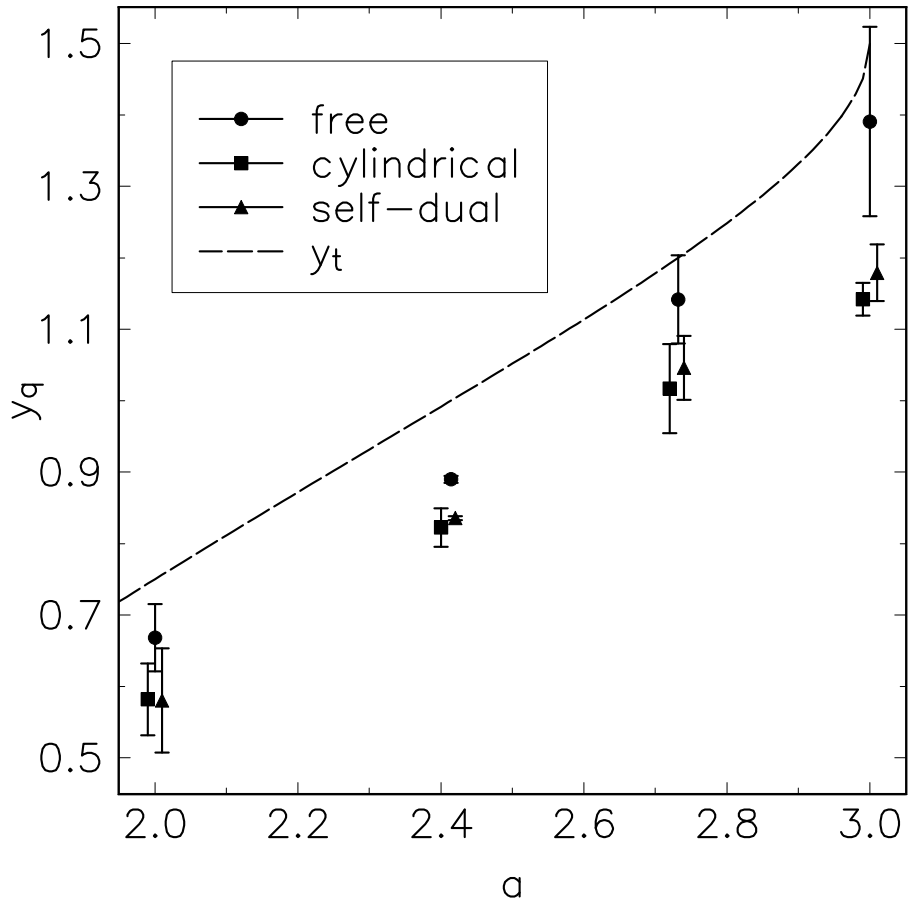}
\caption{The exponent $y_q$ as a function of $a$ for free, cylindrical,
and self-dual boundary conditions. The slight horizontal offset for data
for cylindrical and self-dual boundary conditions is for clarity only.
The long-dashed curve is the thermal exponent $y_t$ by the den Nijs formula.}
\end{figure}

\begin{figure}
\epsfbox{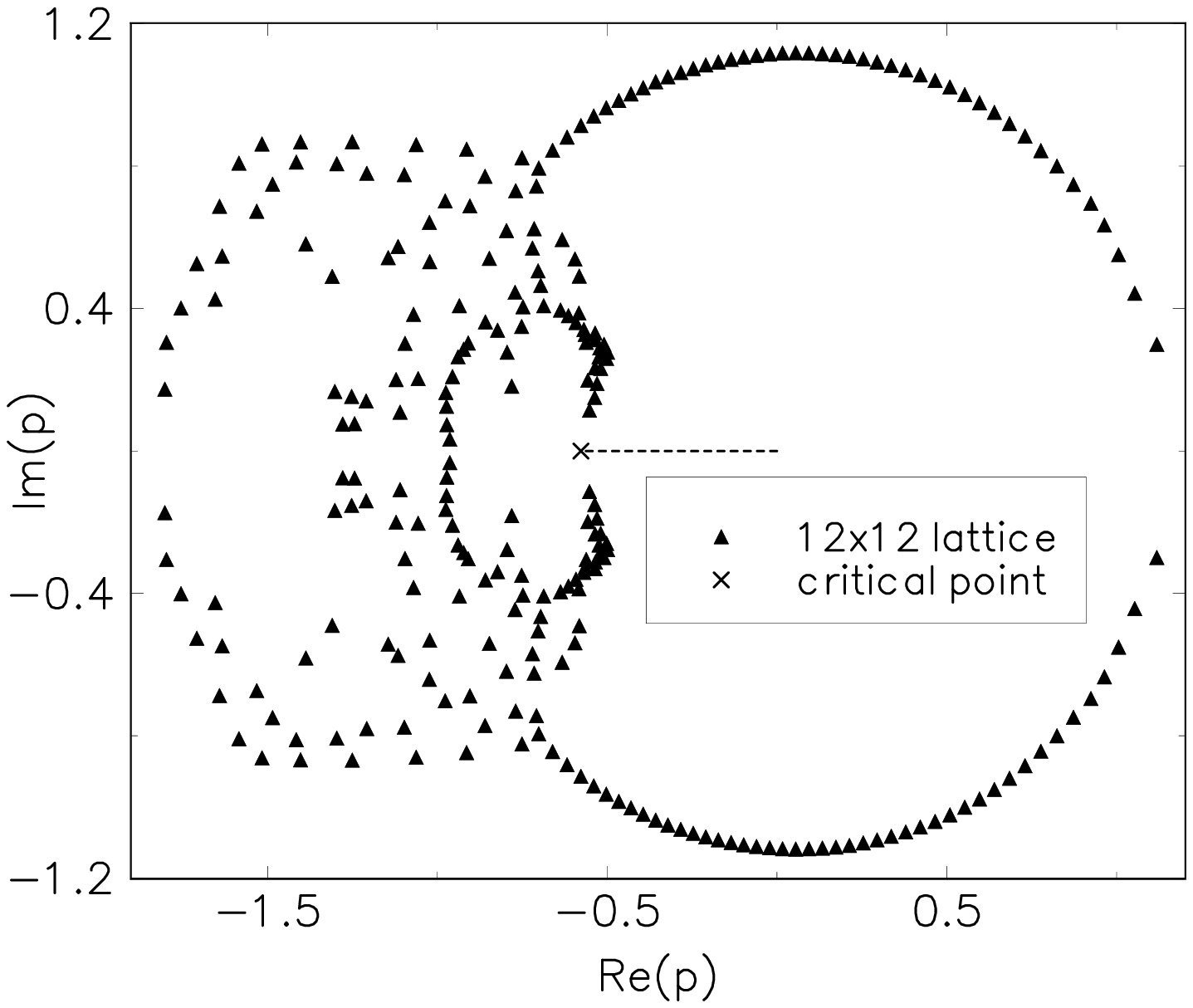}
\caption{Fisher zeros in the complex $p$ plane of the three-state Potts
model on $12\times12$ square lattice with free boundary conditions.
The dotted line is the antiferromagnetic interval.}
\end{figure}

\begin{figure}
\epsfbox{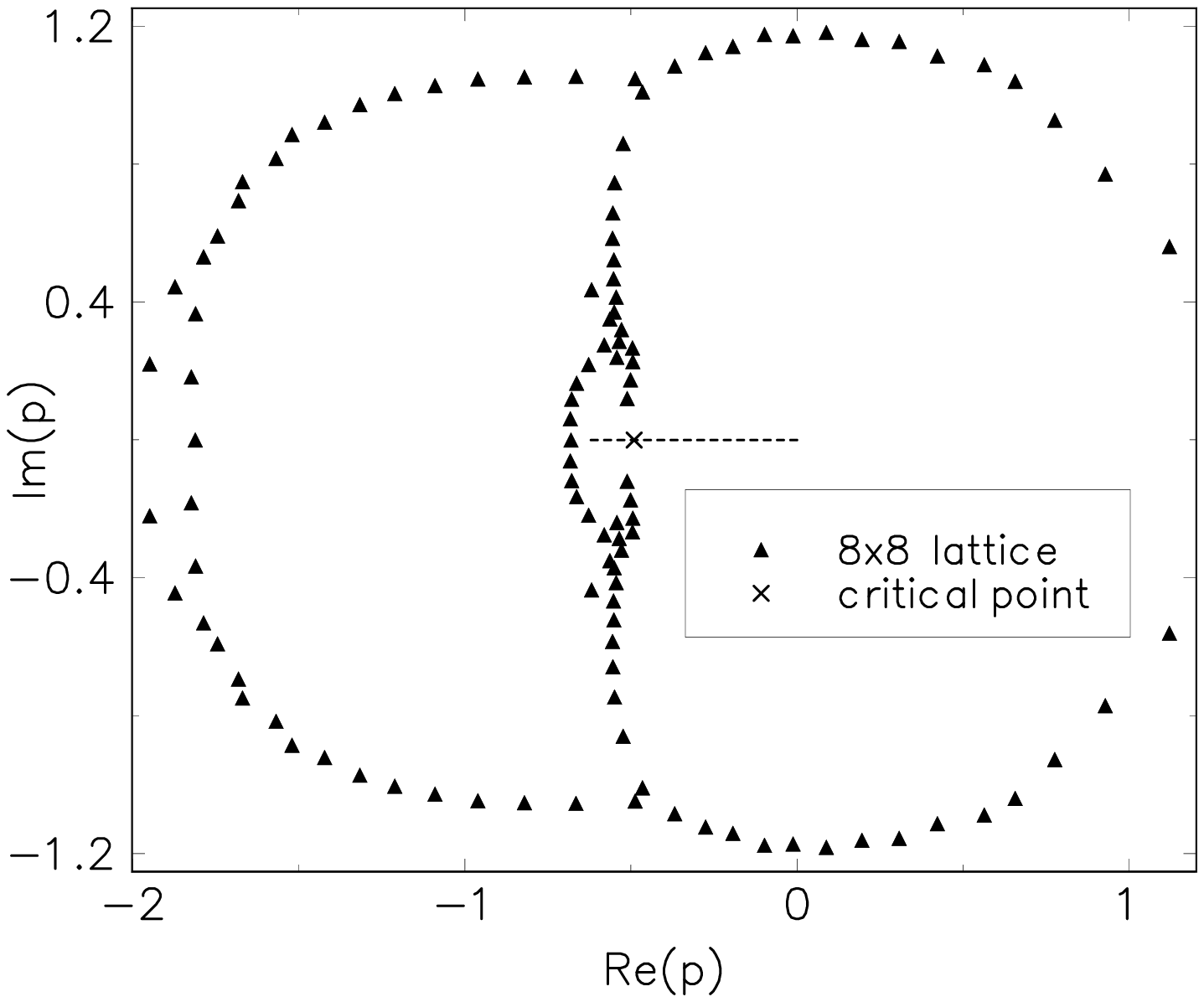}
\caption{Fisher zeros in the complex $p$ plane of the $Q=2.5$ Potts
model on $8\times8$ square lattice with free boundary conditions.
The dotted line shows the antiferromagnetic interval.}
\end{figure}

\begin{figure}
\epsfbox{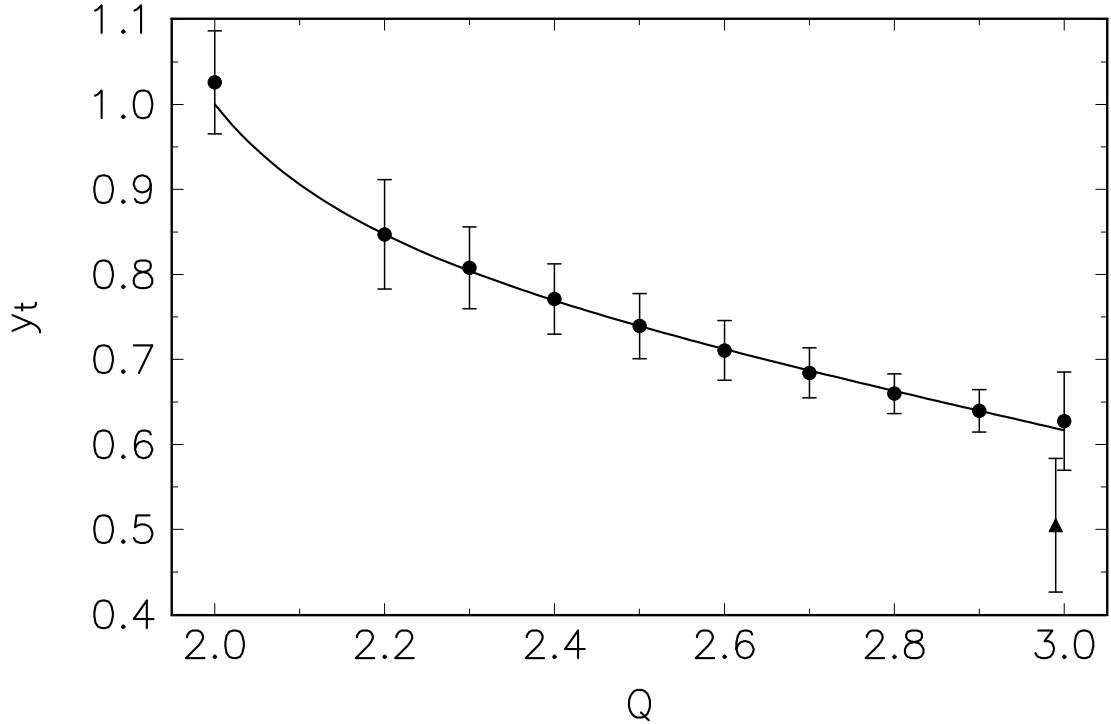}
\caption{The thermal exponents $y_t$ of the $Q$-state Potts antiferromagnets
by the BST estimates (filled circles) from data for $L=3\sim8$ and
free boundary conditions. For $Q=3$ the BST estimate (filled triangle)
from data for $L=3\sim12$ is added and has the slight horizontal
offset for clarity only.}
\end{figure}

\begin{figure}
\epsfbox{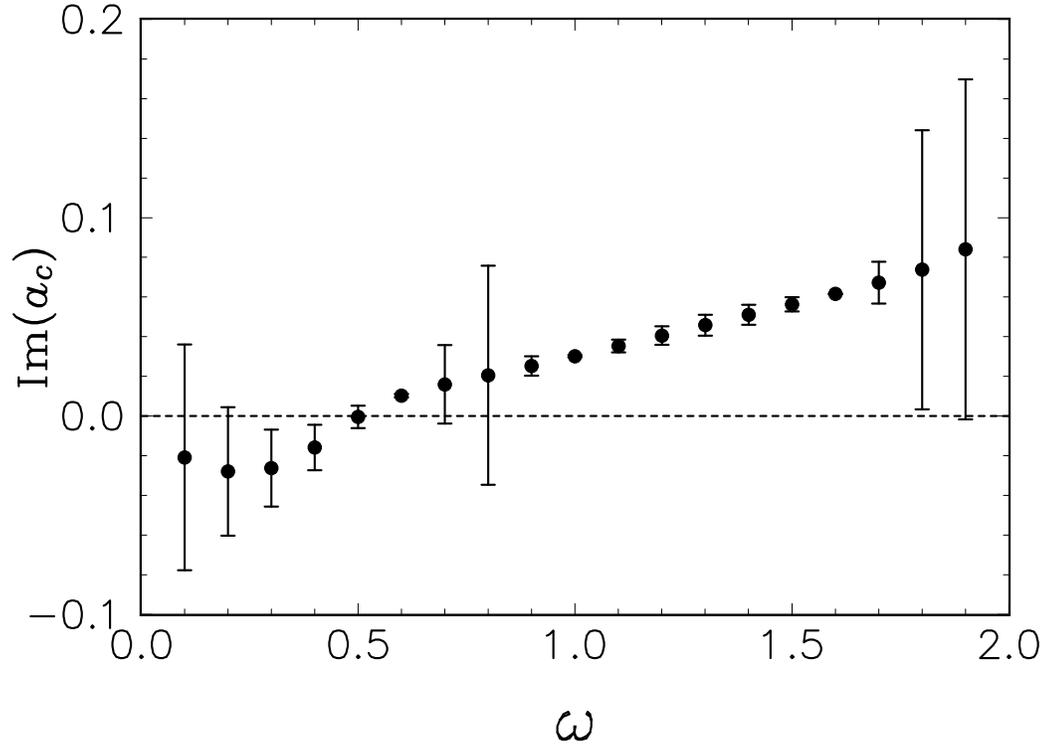}
\caption{BST extrapolation of the imaginary part of the critical point,
${\rm Im}(a_c)$, for the three-state Potts antiferromagnet as a function
of the parameter $\omega$.}
\end{figure}


\end{document}